\documentclass[acmlarge,screen]{acmart}
\AtBeginDocument{%
  \providecommand\BibTeX{{%
    \normalfont B\kern-0.5em{\scshape i\kern-0.25em b}\kern-0.8em\TeX}}}

\setcopyright{acmlicensed}
\acmJournal{PACMHCI}
\acmYear{2024} \acmVolume{9} \acmNumber{CSCW2} \acmMonth{11} \acmPrice{15.00}
\acmPrice{15.00}
\acmISBN{978-1-4503-XXXX-X/18/06}

\usepackage{subfiles}
\usepackage{graphicx}
\usepackage{tcolorbox}
\usepackage{enumitem}
\usepackage{multicol}
\usepackage{colortbl}
\emergencystretch=\maxdimen
\definecolor{Gray5}{gray}{0.95}
\newcommand{\edit}[1]{\textcolor{black}{#1}}

\begin{document}

\title[Epistemic Injustice in CSCW]{Whose Knowledge is Valued?: Epistemic Injustice in CSCW Applications}

\author{Leah Hope Ajmani}
\affiliation{%
  \institution{University of Minnesota}
  \city{Minneapolis}
  \state{Minnesota}
  \country{USA}}
\email{ajman004@umn.edu}
\author{Jasmine C. Foriest}
\affiliation{%
  \institution{Georgia Institute of Technology}
  \city{Atlanta}
  \state{Georgia}
  \country{USA}}
\email{jay@gatech.edu}
\author{Jordan Taylor}
\affiliation{%
  \institution{Carnegie Mellon University}
  \city{Pittsburgh}
  \state{Pennsylvania}
  \country{USA}}
\email{jordant@andrew.cmu.edu}
\author{Kyle Pittman}
\affiliation{%
  \institution{Evergreen State College}
  \city{Olympia}
  \country{USA}}
\email{pittmank@evergreen.edu}
\author{Sarah Gilbert}
\affiliation{%
  \institution{Cornell University}
  \city{Ithaca}
  \country{USA}}
\email{sarah.gilbert@cornell.edu}
\author{Michael Ann Devito}
\affiliation{%
  \institution{Northeastern University}
  \city{Boston}
  \country{USA}}
\email{m.devito@northeastern.edu}

\renewcommand{\shortauthors}{Ajmani, et al.}

\begin{abstract}
Social computing scholars have long known that people do not interact with knowledge in straightforward ways, especially in digital environments. While policies around knowledge are essential for targeting misinformation, they are value-laden; in choosing how to present information, we undermine non-traditional---often non-Western---ways of knowing. \edit{Epistemic injustice is the systemic exclusion of certain people and methods from the knowledge canon. Epistemic injustice chips away at one's testimony and vocabulary until they are stripped of their due right to know and understand. In this paper, we articulate how epistemic injustice in sociotechnical applications leads to material harm. Inspired by a hybrid collaborative autoethnography of 14 CSCW practitioners, we present three cases of epistemic injustice in sociotechnical applications: online transgender healthcare, identity sensemaking on r/bisexual, and Indigenous ways of knowing on r/AskHistorians.} We further explore signature tensions across our autoethnographic materials and relate them to previous CSCW research areas and personal non-technological experiences. We argue that epistemic injustice can serve as a unifying and intersectional lens for CSCW research by surfacing dimensions of epistemic community and power. Finally, we present a call to action of three changes the CSCW community should make to move toward its own goals of research justice. We call for CSCW researchers to center individual experiences, bolster communities, and remediate issues of epistemic power as a means towards epistemic justice. In sum, we recount, synthesize, and propose solutions for the various forms of epistemic injustice that CSCW sites of study---including CSCW itself--propagate.
\end{abstract}

\ccsdesc[500]{Human-centered computing~HCI theory, concepts and models}
\ccsdesc[300]{Human-centered computing~Collaborative and social computing theory, concepts and paradigms}

\keywords{justice, online communities, content moderation, knowledge creation}

\received{January 2024}
\received[revised]{April 2024}
\received[accepted]{May 2024}

\maketitle
\section{Introduction}
Imagine you are a researcher attempting to update your pronouns on your own Wikipedia page. First, you must find a verifiable source with your preferred pronouns. These could be interviews where you express your pronouns but cannot come from your personal webpage or another ``un-verifiable'' source. Then, you must find someone to edit this article for you, as individuals are not allowed to edit their own articles on Wikipedia. Each step in this process distances you as the knower of your identity and risks reinforcing the injustices you already face around your identity offline. 

The above example highlights how \textit{epistemic policies online carry normative repercussions}. Wikipedia has numerous resources on discussing trans and non-binary subjects and editors\footnote{\url{https://en.wikipedia.org/wiki/Wikipedia:Gender\_identity}}\footnote{\url{https://en.wikipedia.org/w/index.php?title=Wikipedia:Manual_of_Style/Gender_identity}}\footnote{ \url{https://en.wikipedia.org/wiki/Wikipedia:WikiProject_LGBT_studies/Guidelines}}. In fact, when Elliot Page came out as a trans man, Wikipedians swiftly updated his page with the correct pronouns. However, the Wikidata community struggled to fit the vast array of queer identities within its existing database structure~\cite{Weathington2023-bx}. This tension speaks to the situated and structural nature of the issue; if left unresolved, stances on verifiability could exacerbate the already rampant misgendering of genderqueer and trans folks. Beyond peer production platforms, online communities are constantly engaging with positions on \textbf{how} knowledge on the internet “ought to be” and \textbf{who} gets due right to be a knower.

 Contemporary philosopher Miranda Fricker describes this phenomenon as \textit{epistemic injustice:} 
\begin{quote}
    ``\textit{[epistemic injustice] is a distinctively epistemic genus of injustice, in which someone is wronged specifically in their capacity as a knower, wronged therefore in a capacity essential to human value."}[p. 1]~\citep{Fricker2007-zh}
\end{quote} Fricker outlines two forms of epistemic injustice: (1) testimonial injustice and (2) hermeneutical injustice. \textit{Testimonial injustice} occurs when one's testimony is not given due credence because of prejudices against the speaker. \textit{Hermeneutical injustice} is less conspicuous, as it is the systemic absence of tools to interpret one's lived experience. Fricker's framework is in harmony with feminist theories of situated knowledge~\cite{Haraway1988-ch}, often used in human-computer interaction (HCI) work~\cite{Muller2022-sj, Campo_Woytuk2020-dn, Suresh2021-lg}. Epistemic injustice expands on these theories to describe the harm caused when stances on knowledge are left unchecked.

While epistemic injustice may seem abstract, it compounds concrete social harms that can cost lives. These consequences are especially salient in online communities, which come with their own rules, norms, and politics~\cite{Reddy2023-zy}. For example, technology-mediated pathways to mental health support, such as suicide hotlines, are often inaccessible. This lack of resources leaves people in crisis without the tools to interpret their own lived experiences~\cite{Pendse2021-yu}. Meanwhile, personal mental health stories on TikTok are valuable knowledge that helps viewers feel less alone~\cite{Milton2023-vu}. However, this information often gets flagged as misinformation. A systemic lack of online health knowledge and community contributes to unconscionable suicide rates in trans populations~\cite{Moody2015-ry, McNeil2017-by, Tebbe2016-lm}. As the amount of information, curation, and subsequent moderation on the internet grows, so does the embeddedness of existing power structures around knowledge. At its worst, epistemic injustice is a violent act---propagating the gatekeeping and discrediting of life-altering knowledge due to the nature of the knower.

Computer Supported Cooperative Work (CSCW) already has a rich history of exploring the policy and practice of knowledge processes within online communities. From peer-production~\cite{Menking2021-ic}, social media groups~\cite{DeVito2021-yv}, citizen science~\cite{Doan2018-ek}, and various other platforms that rely on user contributions~\cite{gilbert2020run, McInnis2021-zu}, we know that sociotechnical applications shape knowledge. However, it is less clear what is at stake when we fail to give this research problem the attention it deserves, particularly as online communities have gotten more visible, used, and complex~\cite{Reddy2023-zy}. We posit that the framework of epistemic injustice provides an accurate and immensely useful description of these issues. Understanding the epistemic injustice within sociotechnical applications is essential to expanding our approach to knowledge practices and subsequent designs. Without the proper tools to do so, we risk unintentionally exacerbating the systemic injustices that our peers with marginalized identities face every day~\cite{Erete2021-xp}.

The existing challenges of epistemic injustice in CSCW applications indicate a need for more intentional identification, investigation, and mitigation of the problem. In this paper, we present a collaborative autoethnography~\cite{Chang2016-bb} to provoke thought and action within CSCW around epistemic injustice. We collected autoethnographic materials from 14 CSCW practitioners in a workshop setting~\cite{Ajmani2023-cp} and collaboratively analyzed them for commonalities and differences. First, we present three firsthand accounts of observed and experienced epistemic injustice from the authors of this paper. Then, we use these cases to articulate several tensions that surfaced during our workshop. We demonstrate that epistemic injustice offers a lens for unifying known but unarticulated phenomena within CSCW. Epistemic injustice also offers an intersectional lens for understanding research: members of the CSCW community are often experiencers, observers, and even unintentional perpetrators through their own systems of research. We leverage the vocabulary of epistemic injustice to motivate calls to action for CSCW practitioners to do right by the people and communities we study.

\section{Conceptual Background: Epistemic Injustice}
Epistemic injustice marries the branch of philosophy centered around knowledge formation (i.e., epistemology) with the phenomena of injustice. To paraphrase Miranda~\citet{Fricker2007-zh}, epistemic injustice is when an individual is stripped of their capacity as a knower in a way that strips them of their due rights and livelihood.

\subsection{\edit{Foundational Concepts}}
\textbf{Epistemology} is the philosophical inquiry of knowledge formation. It is a theoretical discussion that attempts to distinguish individual belief from the collective we call knowledge. In other words, epistemology presumes that within a topic scope, anyone can be a believer, but only certain individuals are knowers. As Kant articulates, ``it is true that all knowledge begins with experience... it is not true that all knowledge arises from experience."~\cite{Kant1934-dy}. 

The epistemic question of how we ``know'' is particularly important in research settings. Oftentimes, the production of scientific knowledge is viewed as an incremental accumulation of facts---a process of getting ever closer to some objective truth \cite{kuhn1997structure}. However, historians, anthropologists, and sociologists suggest that the production of knowledge is intimately social \cite{kuhn1997structure, pinch1984social, latour2013laboratory}. For example, feminist scholar Donna~\citet{Haraway1988-ch} proposed the epistemic theory of ``situated knowledge.''  Concisely,~\citet{Haraway1988-ch} posits all knowledge comes from somewhere and, therefore, is impossible to separate from the context (i.e., a situation) in which it was created. Drawing on these traditions, CSCW and STS scholars have long studied the ways in which ostensibly objective categories embedded in technology are, in fact, partial and highly political \cite{suchman1993categories, d2023data, bowker1999sorting, gitelman2013introduction}.



\textbf{Justice}, within this context, comes from a perspective of due human rights. In Western philosophy, due right is the concept that everyone receives their fair share of resources and opportunity~\cite{Miller1979-cb, Barry2005-pm}. In Indigenous philosophies, the scope of justice expands to doing right by the land or one's ancestors~\cite{Bear2000-vp}. The negative space of justice (i.e., injustice) is the systemic dehumanization of members of society, particularly through an infringement on basic rights~\cite{Young1990-tl, Ajmani2024-qk}. For instance, our introductory example of expressing gender identity on Wikipedia infringes on the basic human right to be the knower of your own gender, pronouns, and overall identity. \edit{In this paper, we focus on occurrences of injustice that are epistemic in nature: they directly remove one's due right to being a knower. Below, we summarize~\citet{Fricker2007-zh}'s articulation of epistemic injustice.}

\subsection{Forms of Epistemic Injustice}
Epistemic injustice serves as an umbrella term for the stripping of one's rights to be a knower, such as the right to be an authority on one's personal experience. In technical settings, epistemic injustice presents as one getting less credence from the hearer(s) than one deserves. An intuitive example of epistemic injustice is mansplaining: to explain something to a woman in a condescending way that assumes she has no knowledge about the topic, simply because she is a woman\footnote{\url{https://www.merriam-webster.com/dictionary/mansplain}}. Like most forms of injustice, epistemic injustice interacts with other societal inequities such as oppression~\cite{Dotson2014-ji} and violence~\cite{Dotson2011-nm}.~\edit{For example, violence is often exercised in relation to the production of knowledge, particularly in academic settings~\cite{Ymous2020-uc}.} In this section, we describe two types of epistemic injustice relevant to this paper: testimonial and hermeneutical. We describe how each form prohibits an individual from their due right to contribute or gain essential knowledge. In the specific context of marginalized communities, epistemic injustice prevents identity groups from accessing necessary resources, such as adequate healthcare information~\cite{online_trans_seeking, redcay_basis_2021, Cohen_Shabot2021-sn} or useful narratives about their own traumatic experiences~\cite{Falbo2022-jy}.

\subsubsection{Testimonial Injustice}
\begin{quote}
    "\textit{Testimonial injustice occurs when prejudice causes a hearer to give a deflated level of credibility to a speaker's word.}" \citep[p. 17]{Fricker2007-zh}
\end{quote}
Testimonial injustice is when a speaker is not given due credence as a knower. This phenomenon is the most intuitive form of epistemic injustice as it can happen at both an individual and societal scale. For example, mansplaining can happen in a one-on-one interaction between a man and a woman. However, it is also systemic as testimonial injustice discounts someone's testimony a priori, typically based on their identity or presentation.

Testimonial injustice removes one's right to contribute to a relevant discourse. Previous work has described this injustice in healthcare settings, where women receive disproportionately worse care than men because their pain levels are not believed by healthcare providers~\cite{Carel2014-ep}. Thus, their treatment ends up ineffective and based on misdiagnoses. In many cases, testimonial injustice is compounded by intersecting marginalized identities. For example, women of color are even less likely to be believed when they report their own symptoms and pain levels in medical settings~\cite{Kidd2017-dl}. In many contexts, such as obstetrics, the discounting of one's testimony costs the lives and livelihoods of victims~\cite{Cohen_Shabot2021-sn}. Testimonial injustice allows harm caused to a specific community, such as women of color, to go unchecked as their stories are seen as non-believable and accusatory. In this way, testimonial injustice \textit{\textbf{is}} systemic harm.

\subsubsection{Hermeneutical Injustice}
\begin{quote}
       \textit{"Hermeneutical injustice occurs at [an early] stage when a gap in collective interpretive resources puts someone at an unfair disadvantage when it comes to making sense of their social experiences."} \citep[p. 147]{Fricker2007-zh}
\end{quote}
Hermeneutical injustice is when one is stripped of resources to gain knowledge or make sense of their own lived experience. Though less intuitive than testimonial injustice, hermeneutical injustice is much more subversive. For example,~\citet{Fricker2007-zh} cites knowledge of sexual assault as a hermeneutical resource that women systemically did not have access to. Historically, when a woman was assaulted, she had no words to describe the experience to her peers, who most likely had similar experiences and could offer support. Previous work has shown that this lack of interpretive resources for traumatic experiences can lead to a lack of support-seeking, feelings of isolation, and even increased suicide rates for trauma survivors~\cite{Meretoja2020-cq}.~\edit{This isolation pushes survivors further into the margins of society~\cite{Galtung1990-kp}. Ultimately, the symbol of a woman who has experienced sexual assault gets annihilated, exacerbating the vicious cycle of injustice~\cite{Tuchman2000-de}}. 

Hermeneutical injustice is not necessarily a lack of sensemaking resources, but also the existence of distorting ones.~\edit{The hermeneutical resource ``sexual assault" came about in the 1970s through collective feminist consciousness raising~\cite{Hogeland2016-gj}. However,~\citet{Falbo2022-jy} notes that the term still carries presumptions of ``some creepy stranger in an alley."} Therefore, familiar faces, such as husbands or bosses, cannot be perpetrators of assault~\cite{Jenkins2017-jy, Jackson2018-lo}. This leaves victims of common but counterintuitive phenomena, such as date rape, left without the appropriate sensemaking resources for their experiences~\cite{Jackson2019-rp}.

\subsection{Epistemic Injustice in CSCW}
Sociotechnical applications are venues for vast amounts of information. For many users, this information guides epistemic processes. For example, hearing a TikTok creator's experience with mental illness can inspire another user to seek a diagnosis and other relevant knowledge~\cite{Milton2023-vu}. Search engines, which are fueled by CSCW applications~\cite{Vincent2021-xl}, are the modern-day library and greatly inform an individual's sense of fact. Other platforms are explicitly positioned as venues of pure factual knowledge, e.g., Wikipedia is the largest peer-produced encyclopedia. These online venues often create blurry lines between information and knowledge. For example, misinformation on social media websites leads to anti-factual narratives~\cite{Efstratiou2022-zh}. Therefore, many CSCW platforms and communities are obligated to create their own systems to determine valid information and knowledge processes.

To that end, platforms have algorithms, policies, and norms that dictate what information is salient~\cite{Gillespie2022-dg}. These moderation techniques are necessary mechanisms for preventing harassment, toxicity, and vandalism~\cite{Jhaver2021-le, Chandrasekharan2022-xb, Geiger2010-ny}. However, moderation in CSCW applications can often be blunt~\cite{Ajmani2023-yk} and disproportionately affect already marginalized identity groups~\cite{Gray2021-wd, hamison_disproportionate_2021, Harris2023-de, Foriest2024-rj}---such as the LGBTQ$+$ community who heavily rely on these online spaces for support~\cite{DeVito2021-yv, DeVito2018-gr, Walker2020-qn}. Moreover, personal experiences on medicalized subjects, such as mental health or trans healthcare, often get flagged as misinformation despite their crucial roles in filling knowledge gaps, and building shared vocabularies of experiences~\cite{Milton2023-vu, Steen2023-gb}. At best, these are undesired side effects of necessary content moderation. At worst, these adverse effects propagate structural injustices~\edit{by furthering the marginalization, fracturing, and silencing of specific identity groups}. 

In the remainder of this paper, we describe three high-stakes CSCW settings where the systemic stripping of resources, credence, and---ultimately---knowledge furthers the oppression of trans, bi, and Indigenous communities. We use epistemic injustice to contextualize the stakes on unchecked stances on knowledge and highlight the potential for future CSCW to remediate and repair this harm. \edit{We build on current calls for CSCW and HCI to coalesce on a specific set of norms and philosophies to guide research paradigms~\cite{Keyes2019-uh, Chivukula2020-gk, Spiel2021-aa}. Specifically, we call for epistemic approaches to considering justice within sociotechnical applications.} Integrating epistemic injustice into CSCW practice not only aligns closer with the field's ideals but also extends the landscape of actionable knowledge even in widely-researched settings,~\edit{such as Reddit, Wikipedia, and our own research-producing community.}

\section{Methods}
In this section, we present an overview of collaborative autoethnography as a method and describe our usage of the method to generate this paper. Our findings are inspired by discussions from a hybrid workshop of 14 CSCW practitioners~\cite{Ajmani2023-cp} using collaborative autoethnography methods and directly draw on the autoethnographic cases of three participants in that workshop who are also authors on this paper. We began our workshop with a primer activity to build common ground. We then prompted participants to write about personal experiences with epistemic injustice. These cases served as autoethnographic materials for a collaborative thematic analysis. Our methodology allows us to hold space for two seemingly paradoxical--yet intersecting--positionalities: (1) researchers who hold the key to the ``knowledge cannon'' of archival research and (2) marginalized folks with lived, traumatic experiences of epistemic injustice. In other words, our methodology hinges on our hybrid subjectivities as both researchers and those who experience epistemic injustice~\cite{abu2008writing}.

\subsection{Methodology: Collaborative Autoethnogaphy}
Authoethnograpy, a self-study method with the researcher as a participant~\cite{Ellis2011-oe}, is a powerful method in CSCW because it necessitates researcher visibility. As~\citet{Rapp2018-im} articulates, increasing researcher visibility allows a researcher's own positionality to appropriately entangle with observations.~\edit{Autoethnography fundamentally seeks to relate a personal experience to cultural experience~\cite{Ellis2011-oe}. In HCI, autoethnography answers deeply personal questions about technology culture. For example, what happens when we reject the cultural norms of mobile phone use~\cite{Lucero2018-gl}? How are we able to personally grapple with being technology experts and still fumbling through authentication ceremonies~\cite{Fassl2023-ur}? Most relevant to this paper,~\citet{Erete2021-xp} used autoethnography to painfully describe their experience as black CSCW researchers during a time of civil unrest. As the lived experience of injustice is often traumatic, autoethnography removes the distance between the expert, experiencer, and knower.}

However, autoethnography has its limitations.~\citet{Lapadat2017-te} notes that, in an effort to center self-study, autoethnographies lack the plurality that other qualitative methods afford~\cite{Braun2006-eg}. In this paper, we are focused on the systemic issue of epistemic injustice. Therefore, it is important to highlight the sweeping and plural effects of injustice. To avoid a mono-vocal work, we adopted~\citet{Chang2016-bb}'s method of collaborative autoethnography. As its name suggests, collaborative autoethnography is the collective analysis of numerous individual experiences. In other words, collaborative ethnography allows researchers to pool reflexive observations and interpret them as a group. This allows for researcher visibility of autoethnography and the plurality of collaborative methods. Specifically, we adopt synchronous collaborative autoethnography, where the analysis is done all together in a conversational hybrid setting.

Given that all of our workshop members are also members of the CSCW community, we present our participant demographics in aggregate to maintain anonymity. In total, we had 14 workshop contributors. We used affirmative consent (opt-in) methods~\cite{Im2021-uu}, such as Google Forms and early manuscript dissemination, to ensure that all members represented in this paper have consented to the submission of this manuscript. Our examples and quotes are from a subset of participants who opted-in after the workshop to their ideas being shared in a paper. Our workshop members included graduate students within CSCW, industry researchers, and professors. Research experience levels ranged from graduate students with a few years of CSCW research experience to professors and industry workers with over a decade of studying knowledge-based communities. Finally, members self-disclosed relevant marginalized identities, including racial, ethnic, and gender identities that have contributed to their experiences of epistemic injustice.

\subsection{Workshop Overview}
\subsubsection{Primer Activity}
In participatory methods, primer activities are useful to set the tone before an experiment and scope the exercise~\cite{Sanders2010-ha}. Our workshop consisted of researchers who study epistemic mechanisms, such as storytelling on social media sites, but not all participants were experts in~\citet{Fricker2007-zh}’s framework of epistemic injustice. To solve this broad background differential, we began with a primer activity designed to share and unpack our own interpretations of the term. 

We introduced a one-line definition of epistemic injustice: \textit{“The systemic exclusion of voices from contributing to knowledge.”} We then asked participants to take three post-it notes and complete the following steps:
\begin{itemize}
    \item[] Post-it note 1: Your interpretation of epistemic injustice
    \item[] Post-it note 2: An iteration of post-it note 1
    \item[] Post-it note 3: An iteration of post-it note 2
\end{itemize}
We then had participants pair off to share their interpretations and note down common concepts.
\begin{figure}
    \centering
    \includegraphics[width=0.9\textwidth]{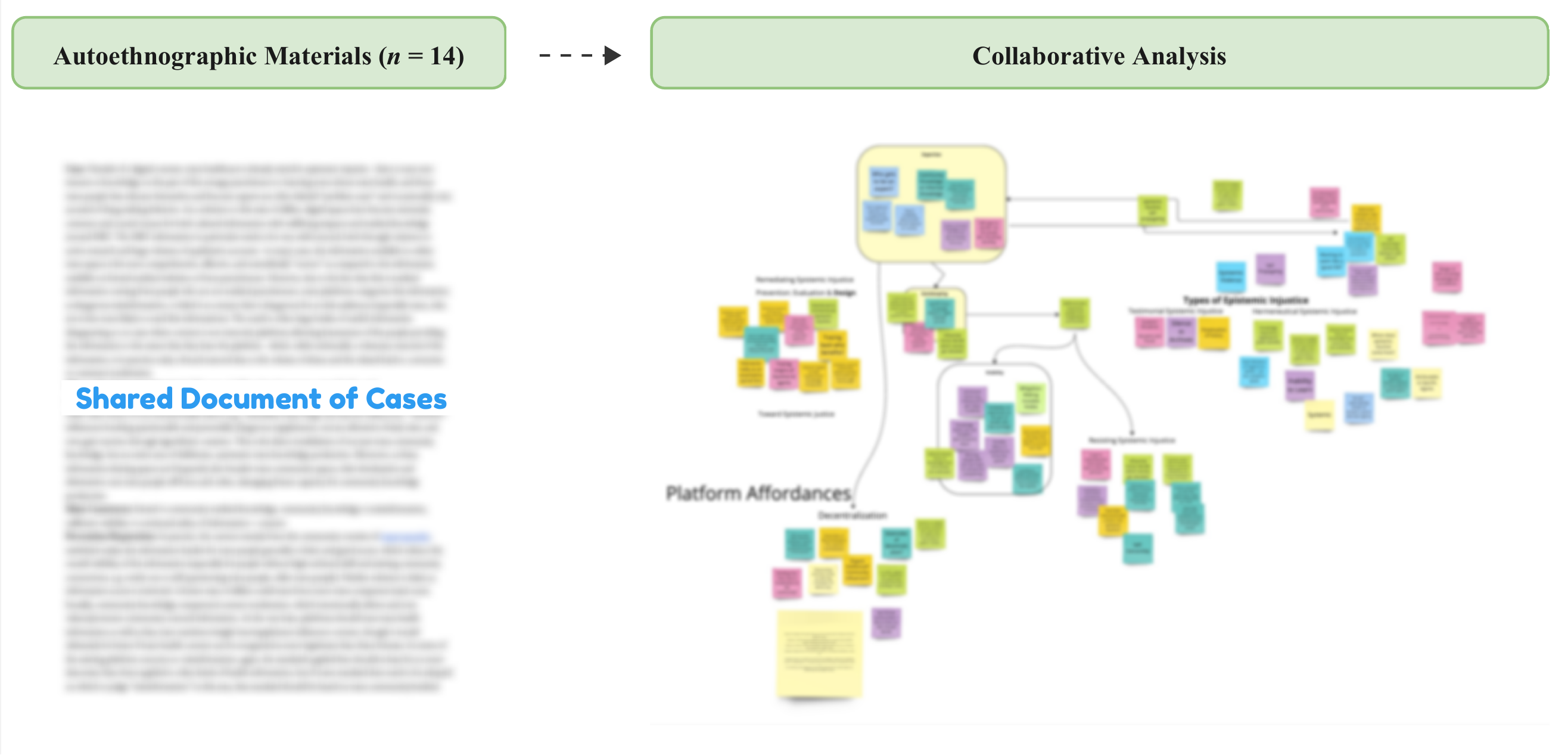}
    \caption{Analysis of autoethnography materials in an online whiteboard with collaborative document editors. Materials are blurred to maintain anonymity.}
    \Description{}
    \label{fig:materials-overview}
\end{figure}

\subsubsection{Autoethnographic Materials}
For the autoethnographic portion of our workshop, we prompted participants to reflect on a case study of epistemic injustice in a shared document with the following instructions:
\begin{quote}
    \textit{This document aims to highlight cases (both anecdotal and empirical) where epistemic injustice manifests in sociotechnical systems. We hope this document demonstrates the breadth of platforms and design patterns contributing to marginalized communities' systemic silencing. Furthermore, we hope to note ways to prevent these outcomes or redress the harms caused. }
\end{quote}
A workshop facilitator also verbally introduced the activity to the hybrid room of participants. Workshop members had approximately twenty minutes to individually write about cases of epistemic injustice that they have experienced or observed. While members were not told this was a silent activity, the reflective nature of the exercise paused conversation among the group. As a starting point, we provided the following template for each case
\begin{itemize}
\item \textbf{Title:} 
\item \textbf{Author: }
\item \textbf{Case: }
\item \textbf{Why is this Epistemic Injustice?:}
\item \textbf{Main Constructs:}
\item \textbf{Prevention/Reparation:}
\end{itemize}
Workshop members were encouraged to use the guiding template however best served them.

\subsubsection{Collaborative Analysis}
After all the cases were written, we ran a hybrid, collaborative thematic analysis of these cases. We shared a Miro board with all participants and went through case by case. For each case, we opened up a discussion where participants noted important features of the individual case. Contributors noted features verbally in discussion and with Google Doc comments on the shared cases document. This joint verbal and text method was particularly helpful in our hybrid setting, where it can be difficult to capture virtual and in-person voices. A workshop organizer served as a scribe and documented each feature--from both the ongoing discussion and document comments--as a virtual post-it note. Participants were welcome to move post-it notes around and add contextual labels on the shared board as they saw fit. After every two cases, we paused for five minutes so that participants could continue to contribute, cull, and cluster post-it notes. This method of having a shared, visible resource that all participants can manipulate is inspired by social translucence and has been used in previous HCI collaborative autoethnographies~\cite{Bala2023-mf}.

\edit{In this paper, we present several components of our workshop. First, we present three firsthand testimonies of experienced and observed epistemic injustice from the authors of this paper. These are edited versions of the autoethnographic cases generated in our workshop. Second, we present several tensions that surfaced during our workshop and relate to broader phenomena in CSCW and HCI. All quotes from our workshop were approved by their respective participant.}

\section{Autoethnographic Testimonials}
\begin{figure}[H]
\begin{tcolorbox}[boxrule=0pt, frame empty]
\begin{multicols}{2}
\begin{itemize}
\item Trans Health information online
\item PhD Application Information online
\item r/AskHistorians
\item Facebook groups
\item TikTok
\item r/Bisexual
\item Peer Review Process
\item Policing
\item Environmental Activism online
\item LLMs
\item Search Engines
\item Recommendation Systems
\item Wikipedia
\item Social Media Platforms
\end{itemize}
\end{multicols}
\end{tcolorbox}
\caption{List of venues mentioned in autoethnographic materials}
\label{fig:venues}
\end{figure}
 Given the broad prompt, not all of the venues mentioned in our workshop are traditional CSCW applications (Figure~\ref{fig:venues}). However, many of the online venues mentioned are established and widely-researched sociotechnical spaces, such as Reddit and Wikipedia. Participants noted that epistemic injustice does not rear its head simply when a human is interacting with a computer (i.e., in individualistic HCI settings). We posit that CSCW applications are ripe venues for epistemic injustice as technical systems often mediate community information, norms, and shared knowledge. 

In this section, we present three testimonies of experienced and observed epistemic injustice within CSCW and related spaces.~\edit{We use these three stories (1) as data that we synthesize into a larger analysis of epistemic injustice in CSCW (Section~\ref{sec:tensions}) and (2) an illustration of the power of giving testimony its due place in research. First-person research sometimes ventures into the overly scientific and can be constrained by traditional research expectations~\cite{Lapadat2017-te}. We encourage readers to consider these testimonies as the result of researchers writing their own stories and then systematically analyzing and iterating upon their articulations. The authors of these testimonies chose to relate their personal stories to related research because many of us are CSCW researchers ourselves. In other words, these testimonies lay at the intersection of us as knowledgeable academics with traditional credentials, CSCW practitioners who perpetrate epistemic injustice through our systems or research, and individuals who move through the world with visible and invisible marginalized identities.}

\subsection{Testimonial Injustice in Online Trans Health Information}
\textbf{Positionality.} As a transgender woman and a computer scientist who has expertise in online communities, content moderation, and information seeking, I am deeply invested in the promotion of well-sourced transgender health and wellbeing information via digital mediums. Here, I pull from both my personal experience as a transgender woman and from my own research and expertise, as well as the published work of other queer and transgender scientists. \textit{(Michael Ann Devito)}

\textbf{Case.} Outside of a digital context, healthcare for transgender people is already mired in injustice: lack of practitioner knowledge and interest, significant legal barriers \cite{BAUER2009348, redcay_basis_2021}, and outright discrimination within healthcare settings \cite{kcomt_profound_2019} lead to high levels of negative physical and mental health outcomes for transgender people \cite{drabish_health_2022}. Absent these traditional pathways to healthcare and health information, digital spaces have become extremely common and crucial venues for both transgender cultural information with wellbeing impacts and practical medical knowledge \cite{online_trans_seeking}, especially around gender affirming hormone treatments (GAHT)~\cite{edenfield_always_2019}. The GAHT information, in particular, tends to be very well-sourced, both through citations to active research and large volumes of qualitative accounts. In many cases, the information available in online trans spaces is far more up-to-date, comprehensive, and effective as compared to the information available on formal medical websites or from practitioners \cite{edenfield_queering_2019, alexander_health_2021}. Essentially, trans people themselves, especially those trans people who choose to assume an educational role within online trans spaces, have become the functional authority on transgender health information at an on-the-ground level, even building comprehensive communities around this kind of information sharing \cite{edenfield_tactical_2019}. 

Despite the large amount of effort trans educators put into securing, updating, and maintaining such spaces, including the day-to-day work of making sure these spaces are safe and well-curated for all trans people (including minors) \cite{DeVito2022-ee}, this crucial online trans health information is at constant risk of erasure. Trans-related content, including trans health content, is disproportionately removed by content moderation systems. In some cases, this appears to be simply because it is linked to transgender identity and connected to the accounts of transgender people, while in other cases content moderation systems are triggered by certain types of content (e.g., surgical images) \cite{hamison_disproportionate_2021, online_trans_seeking}. Due to the fact that this is medically-related information coming from people who are not medical practitioners, some platforms categorize this information as dangerous misinformation, or label it as content that is dangerous for at-risk audiences \cite{Milton2023-vu, online_trans_seeking} (especially teens, who are in fact most likely to need this information). Additionally, trans health and wellbeing information also falls victim to removal due to sustained campaigns of false reporting and online harassment towards trans people \cite{DeVito2022-ee}. The end result is often large bodies of useful trans health and wellbeing information disappearing or, in cases where content is not removed, platforms allowing harassment of the people providing the information to the extent that they leave the platform. While leaving is technically a voluntary removal of the information, it is in practice truly a forced removal due to the volume of abuse, the personal impact of the abuse on trans educators, and the related heavy, uncompensated workload around moderation and correction in comment sections and other interactive forums \cite{DeVito2022-ee}. Moreover, my personal experience and the experience of many trans people I have interviewed suggests that even when we can find online trans health information, attempting to share this information with medical practitioners can sometimes result in another layer of disbelief and dismissal from said practitioners, with those trans people that educate themselves and become functional experts even being labeled “problem cases” or "combative" in their formal medical records simply for attempting to fill a practitioner knowledge gap. In turn, this heavily contributes to a continued state of poor health outcomes and inadequate access to care for trans people. 

\textbf{Why is this Epistemic Injustice?} This state of affairs directly removes scientifically- and community-validated, high-quality information from the public sphere, and does so specifically for trans people, while other forms of non-formal health advice, such as supplement ads, are allowed to freely exist, and even gain traction through algorithmic curation. This is the direct invalidation of not just trans community knowledge, but an entire area of deliberate, systematic trans knowledge production. Moreover, as these information-sharing spaces are frequently also broader trans community spaces, their devaluation and elimination cuts trans people off from each other, damaging future capacity for community knowledge production. This constitutes testimonial injustice, the form of epistemic injustice where a person or community's knowledge is devalued due to their identity or positionality \cite{Fricker2007-zh}. Trans people have effectively been forced to curate and produce our own medical expertise through no fault of our own - indeed, many if not most of us would much rather interact with a medical system where the doctors truly are the experts in our care, but this is simply not the state of the world. And yet, despite the fact that we have put so much collective effort into building a base of knowledge we should have never been required to build in the first places, the fact that this information comes from us, from people with our identities, marks it as knowledge that practitioners and platforms mechanisms alike are seemingly determined to ignore, erase, and invalidate. We are not allowed to be the experts about ourselves, our experiences, or our health needs, simply because of who we are and the negative social and professional stigma around trans identity. Intentionally or not, overzealous platform moderation mechanisms and poorly scoped or enforced platform policies directly contribute to continued epistemic injustice in this area, functionally erasing and invalidating our testimony in favor of a continued overall lack of knowledge. This is the depth of the injustice here: practitioners and platforms alike create a situation where a complete lack of knowledge is preferred to believing and acting on the testimony of trans people.

\textbf{Prevention \& Reparation.} In practice, the current remedy from the community consists of steganography, which makes the information harder for trans people generally to find, and gated access, which reduces the overall visibility of the information (especially for people without high technical skill and existing community connections, e.g. newly out or still questioning trans people, older trans people). Neither solution is ideal, as information access is restricted. A better state of affairs could stem from more trans-competent (and, more broadly, community-knowledge-competent) content moderation, which intentionally allows and even values/promotes community sourced information. In terms of the existing platform concerns regarding potential misinformation, the standards applied to trans health information should at least be no more draconian than those applied to other kinds of health information, but if some standard does need to be adopted on which to judge “misinformation” in this area, that standard should be based on collaborations between the trans community and medical establishment. For example, the TPATH standards are created by trans academic health professionals instead of the WPATH standards, which are mostly promulgated by cisgender doctors. Moreover, undue targeting of such content for creator abuse should be viewed by platforms the same as targeting a doctor or health institution would be.

\subsection{Researching Epistemic Injustice in a Bisexual Online Community}

\textbf{Positionality.} In this case, I reflect on my experience as a bisexual social computing researcher studying bisexual online communities. I provide background on bi-erasure: an injustice that the bi community faces both from within the LGBTQ$+$ community and from outside. I speak to my own tensions and concerns of being a member researcher within the bisexual community. \textit{(Jordan Taylor)}

\textbf{Case. } The meaning of the language one uses to describe sexuality can differ from person to person. However, bisexuality is often understood as an ``umbrella'' term for those who are neither exclusively straight nor gay \cite{flanders2017under}, such as pansexuals or omnisexuals. Bisexual identification is widespread. In a 2023 Ipsos poll of adults in 30 countries, 5\% of respondents identified as bisexual, pansexual, or omnisexual \cite{Jackson_2023}. Despite its prevalence, bisexuality has been largely underrepresented and misrepresented in media \cite{glaad_2021, magrath2017bisexual} and law \cite{yoshino1999epistemic, marcus2018global}. The legal scholar Kenji~\citet{yoshino1999epistemic} coined the term ``bisexual erasure'' to make sense of this underrepresentation. Yoshino argues this erasure is not incidental but rather the result of an unspoken agreement between straight and gay people who, for instance, both have shared interests in stabilizing sexual orientation \cite{yoshino1999epistemic}. Similarly,~\citet{walker_more_2020} found that bi people can experience erasure and invalidation from both those outside and within the LGBTQ+ community, particularly in online spaces.

My research examines the social construction of bisexuality in online communities. Specifically, I recently studied the bi-specific subreddit---r/bisexual \cite{r_bisexual}. Reddit is a social media platform with community-moderated forums, known as subreddits, centered on specific topics. In alignment with previous scholarship on bisexuality, members of r/bisexual believed bisexuality was often misunderstood and underrepresented in the public sphere, causing some to struggle to understand themselves or be understood by others. To fill this knowledge gap, I found members of r/bisexual constructing a system of meaning to understand how to be and do bisexuality, such as inventing new and nuanced language.

I was initially motivated to study r/bisexual due to the lack of HCI research focusing on the particular experiences of bi people, seeming to perpetuate the epistemic injustice of bi erasure. While I often encountered HCI research about LGBTQ+ people, there was a clear gap in bi-specific HCI research. I also knew from my personal experiences that bi people face unique challenges both online and offline, such as exclusions from both straight and gay spaces and technologies. As a personal example, I often struggled with context collapse \cite{marwick2011tweet} on the dating app Tinder because designers expected me to use the same profile when presenting myself to both straight and queer audiences. I worried that research focused on ``LGBTQ+ people'' or ``gay and bisexual men'' likely would not be able to capture this specific experience. I hoped my study of r/bisexual would provide an opening for more work on bisexuality in HCI.

At the same time, I also worry about perpetuating epistemic injustice by misrepresenting r/bisexual and other bi-specific sites of study. While I identify as bi, I had not participated in r/bisexual or related spaces prior to my research career. As such, my understanding of bisexuality differs from those of the r/bisexual community. Even within r/bisexual, there are a plethora of different interpretations and understandings of one's sexuality. However, in speaking on the ``bi community,'' a reader might flatten this heterogeneity as the experience of "all Bisexual People." My concern with misrepresentation in bi-specific HCI research is further amplified by my relative position of power. While there is always a power differential between the researcher and the researched, I am cautious that readers may view my writing about other bisexual people as especially authoritative because I am bisexual. In light of these realities, I worry my research could exacerbate the epistemic injustices I aim to combat.

\textbf{Why is this Epistemic Injustice?} The lack of knowledge regarding bisexuality in the public sphere caused members of r/bisexual to poorly understand themselves or be understood by others, constituting a hermeneutical injustice. The knowledge gap parallels Fricker's example of women struggling to understand shared experiences prior to the advent of the term "sexual assault" within feminist consciousness raising sessions \cite{Fricker2007-zh}. Much like the counter-knowledge of sexual assault, members use r/bisexual to partially overcome their shared knowledge gap. At the same time, my own research could exacerbate this hermeneutical injustice by projecting my individual experiences and beliefs as a bisexual person onto those I was studying.
    
\textbf{Prevention \& Reparation.} Members used r/bisexual to help classify bisexuality or understand what it means to \textit{be} bisexual in the face of common discourses in the public sphere that erase or invalidate bisexual identity. In doing so, members were able to mitigate shared epistemic injustice. For example, designers of r/bisexual advocate for an anti-essentialist, anti-categorical understanding of bisexuality, which they integrated into a post in the subreddit's sidebar. Members also circulated counternarratives addressing who can "be" bisexual. For instance, there were many discussions of the form "you're still bi if [disqualifying experience]." These disqualifying factors included being "older," a man, or in a "mf"  relationship (i.e., one with a man and a woman). Furthermore, members sometimes struggled to square their lived experiences with a social expectation that one has a stable sexual orientation because some experienced fluctuations in their attraction to different genders over time. To make sense of fluctuations in attraction, those in r/bisexual invented language to describe it: the "bi-cycle." The "bi-cycle" was mentioned by both our participants and embedded into the design of r/bisexual via the "Bi-Cycle/Questioning" post flair. Across these different instances of knowledge building, members construct a system of meaning for \textit{being} bisexual that counters dominant discourses. 

In my writing, I sought to address my personal concerns about exacerbating epistemic injustice in research through various strategies. Through positionality statements, I proactively resisted the notion that a single bisexual person has the expertise to speak on behalf of all Bisexual People. I tried to discuss specific communities, such as r/bisexual, as a social world with a particular system of meaning for understanding bisexuality rather than a proxy from which to characterize the experiences of Bisexual People broadly. Through this particularity, I sought to avoid that which the cultural anthropologist Clifford Geertz cautions as "regarding a remote locality as the world in a teacup or as the sociological equivalent of a cloud chamber" \cite{geertz2008thick}.

\subsection {Fighting Epistemic Injustice: Supporting Equity and Navigating Moderation Labor in r/AskHistorians} 
\textbf{Positionality.} 
ta’c léehyn. ‘íin wen’íikt wées Kyle. I am a nimíipuu (Nez Perce) descendant, moderator of r/AskHistorians and r/IndianCountry, and a professor of Native American \& Indigenous Studies. I was raised on a reservation and my professional career has revolved around working in and with Indigenous communities. It is from these perspectives that I’ve contributed to this section. With the potential to reach millions of users, online communities are often at the forefront of manifesting knowledge in ways that impact the general public in how we see, read, and understand the past and even the contemporary world around us. Due to centuries of colonial violence and assimilation attempts, Indigenous knowledge has struggled to be seen as valid in the eyes of settler society and faces many challenges when being introduced into primarily non-Indigenous spaces. Here, I draw upon these experiences with the goal of asserting epistemic justice by creating inclusive moderation policies and advising on approaches that stem the denial of Indigenous epistemologies. \textit{(Kyle Pittman)}

I am an r/AskHistorians moderator and an expert on community content moderation and informal learning. My contributions to this section are drawn from my observations as a moderator and are framed by this expertise. While I see the potential for online communities to provide unique opportunities to learn and share knowledge, my work has also shown how people and sociotechical systems shape participation, and therefore who is included (and excluded) in learning and teaching. It is also important to note that, as a white person whose ancestors were and as someone who continues to benefit from white settler colonialism, my contributions are not drawn from personal experience with the epistemic injustice we describe. Among the mod team, I am part of the non-Indigenous majority that is primarily familiar with Western epistemologies. \textit{(Sarah Gilbert)}

\textbf{Case.} r/AskHistorians, a community on Reddit, has a mission: public history. To that end, the community uses a question-and-answer format, coupled with an extensive set of rules for providing answers, to ensure that while anyone can ask and provide information, the answers they receive will be trustworthy. Reddit supports pseudonymous participation, which means that traditional markers of expertise, such as working in a history-related field, publishing historical research, or having a degree in history are unavailable. Supporting contributions from “lay-historians” is an integral part of r/AskHistorians mission, as it helps promote a more inclusive environment by enabling contributions from people who are typically left out of history discussions. In lieu of these traditional markers moderators have developed and routinely enforce rules that focus on the content of responses rather than characteristics of the answerers, which means that the verifiability of information is important. While sources are not required up front, they are required upon request and anecdotes are not allowed. 

However, rules that center verifiability create challenges when dealing with unrecorded history. This, in turn, makes it challenging for Indigenous community members to share their expertise. While there are markers for methods used in Indigenous ways of knowing, such as recognized cultural knowledge carriers (typically those identified as “elders”), attachment of knowledge to place-based geographic locales (sometimes referred to as traditional ecological knowledge), and a robust practice of maintaining oral traditions (generational transmission of knowledge with rigid protocols) \cite{karson2014wiyaxayxt,chilisa2019indigenous}, these are not accounted for in the rules and most members of the mod team are not versed in Indigenous epistemologies and methods. This puts undue effort on Indigenous moderators who have to first, identify answers that use these markers of knowledge; second, negotiate with others on the mod team about decisions regarding these answers; and third, have to educate the mostly non-Indigenous team on Indigenous methods and how to identify them. It also requires them to be available to address conversations when they arise, which isn’t always the case among a group of unpaid volunteers.

While the mod team would ideally like to account for and allow Indigenous knowledge, it is not always easy to recognize; sometimes, it is even difficult for moderators who are accustomed to Western epistemic traditions to accept the validity of expressions of Indigenous knowledge that do not at least partially rely on conventional methodologies. For example, in one instance an answer was removed when a question-answerer used an elder in their community as a source. Typically, answers that reference someone the answerer knows as a source of information are removed for violating the rule disallowing anecdotes. However, knowledge passed down from elders in their communities is a common and reliable form of information among many Indigenous communities—a point that was not brought up until later, after the answer in question had been removed, effectively silencing the user.  

Removing rule-violating comments is a net positive for the community. Because Reddit’s technological system rewards comments that are short (these tend to be seen first, receive upvotes, and are promoted), removing such comments creates space for longer, more comprehensive answers. Similarly, by removing comments that are bigoted or abusive, r/AskHistorians creates a safer and more inclusive space for users \cite{Gilbert2023-pg, gilbert2020run}. However, in some cases these removals can have the opposite impact from what moderators intend. When answers drawing on Indigenous methods are removed, space is instead created for answers that reflect white, Western, approaches to knowledge. In turn, these answers are rewarded with upvotes; when the upvoting system propagates the information in the answers, Western ways of knowing are reinforced at the expense of Indigenous epistemologies and community members. 

While recognizing individual cases may be challenging for moderators not familiar with Indigenous ways of knowing, the moderation team does generally recognize that this is a problem and has expressed a desire to change it. Not surprisingly, revising the policy so that it is clear to users and to moderators takes work, particularly because the stakes are high—creating, enforcing, and communicating clear and consistent rules is vital to the health of online communities \cite{jhaver2019did, jhaver2019does, koshy2023measuring}. It also requires expertise. While AskHistorians has several members of the mod team who are versed in oral history as a research method and who specifically identify as Indigenous persons with relevant cultural experience regarding oral traditions, crafting a policy to accommodate for Indigenous epistemologies requires experts who can meld these approaches to history in a way that preserves the mission of the community. As such, only a select few on the mod team are able to leverage the knowledge and perspectives needed for such a task. In 2019, two moderators began drafting a potential framework that would allow history that has been passed down orally in some cases. This involved establishing clear definitions of related terms and concepts such as oral tradition and oral history, outlining best practices for using oral methods to convey credible information, and listing potential challenges for configuring these parts into actionable rules of enforcement.

Unfortunately, this project has stalled and none of the other members of the moderation team have the expertise required to take over. The most straightforward method of reparation itself generates a form of racialized labor in which certain members of the mod team may become responsible for contributing a disproportionate amount of time and energy to crafting policies or launching initiatives. In this case, the drafted policy for oral information was also primarily authored by a non-Indigenous moderator, resulting in several portions reflecting a cultural bias that still precludes some forms of Indigenous methods. In one passage, the use of “personal or close family experiences” are barred, citing information stemming from familial relations such as "mothers/uncles" as being disallowed. While this restriction may be reasonable from the perspective of disallowing anecdotes, it ignores the kinship and familial structures utilized by Indigenous communities that result in them often being highly interrelated. For example, while the policy could allow an answerer to cite an elder of their community as the basis for their commentary, this caveat would subsequently disqualify their answer should the elder be a “close” relative. Furthermore, the proximity of relatives would also need to be expanded as the concepts of nuclear and extended families differ among Indigenous Peoples. A Western understanding of a familial relationship may render a relative being classified as a distant cousin whereas an Indigenous understanding may classify that same relative as being an uncle/aunt, a grandparent, or a sibling; these cultural constructions can be difficult to capture in a succinct moderation policy and demonstrate the problems with approaches that do not adequately include the voices of those that may be the subject of said policies.

In other instances, the drafted policy required the answerer to be versed in the scholarship surrounding the subject they were commenting on outside of the oral information they were providing and stressed the contextualization of the oral information within Western standards of epistemic veracity. Answerers who cited oral sources would need to “know the scholarship and be reasonably confident there is not a written source that can supply the same [information].” Additionally, should a request for sources be made, the answerer would need to “include related written sources” and potentially direct others to entities that could provide further information about the “oral tradition/rituals.” While AskHistorians has rules and standards that parallel these requirements, such as expecting familiarity with related scholarship or providing sources upon request, these outlined expectations neglect the history of abuse of Indigenous knowledge and demand that Indigenous ways of knowing conform to Western research norms in order to be considered legitimate. Many oral traditions were stolen from Indigenous communities throughout the 19th and 20th Centuries by being unethically recorded by researchers who then went on to publish this information\cite{smith2021decolonizing}. Under circumstances like this, it would be unreasonable to expect an answerer who may be Indigenous to be familiar with what could be considered an illegitimate source to them, particularly if that source distorts the meaning of the information, and to use it to legitimize their ancestral knowledge. Similarly, forcing answerers to ensure that their oral traditions are backed by written sources undermines the primary purpose of this policy by making oral information incidental and easily dismissable by a request for alternative sources, supporting the cultural supremacy of written sources and further diminishing the validity of oral traditions.

\textbf{Prevention \& Reparation.} In situations where the voices of marginalized persons may be few in number, such as among the ranks of subreddit communities and mod teams, a disparity in labor may be anticipated or even considered inevitable when authentic perspectives are deemed necessary to achieve accurate representation; this is the reality for marginalized peoples when trying to challenge dominant systems that intentionally or inadvertently exclude them. To balance this, it is important for members of a moderation team, particularly those with dominant normative identities, to regularly review their efforts and actions for implicit biases or overtly harmful practices so as to create more inclusive environments.  Furthermore, it is necessary to create an online environment that is not only welcoming for Indigenous persons, but that is also actively combatting Western epistemic hegemony. This can be done by honoring Indigenous ways of knowing, thinking, and being through consultation efforts with Tribal Nations, Indigenous scholars, and Native persons embedded in their communities; cultivating a friendly intellectual space that recognizes the multiplicity of thought and the legitimacy of non-Western philosophical traditions; and developing community-based moderation practices and policies that provide avenues for effectively incorporating different realms of thought into the discourse rather than punitively dismissing them as violations of the rules. These efforts contribute to the normalizing of Indigenous knowledge and work to prevent further injustices from being committed.

\section{Navigating the Tensions of Epistemic Injustice in a CSCW Context}~\label{sec:tensions}
Inspired by~\citet{Bala2023-mf}, collaborative autoethnography surfaces differences among autoethnographic materials. The above vignettes highlight the multifariousness of epistemic injustice---bearing the complexities of knowledge curation in sociotechnical settings. Here, we draw from both our highlighted cases and the larger discussions that took place in the workshop to synthesize tensions and differences among our autoethnographic materials. Throughout our workshop, certain ideas were framed as double-edged swords: pillars of epistemic injustice that have both good and bad consequences. In this section, we highlight common tensions illustrated in the above vignettes and place them in conversation with other relevant cases, such as popular CSCW sites of study and personal experiences with epistemic injustice. We suggest using epistemic injustice as a lens is about understanding where certain ideals, such as expertise, are appropriate versus where they are weaponized to further marginalization.
\subsection{Invisibility}
Invisibility is a nuanced construct within epistemic injustice as it is both a \textit{product of} and a \textit{contributor to} epistemic injustice. Specifically in online communities, epistemic injustice causes certain stories to never appear within the community. As discussed in the case of Indigenous knowledge within r/AskHistorians, invisibility can be hard to detect and even harder to remediate. 

\begin{figure}[H]
    \centering
    \includegraphics[width=0.5\textwidth]{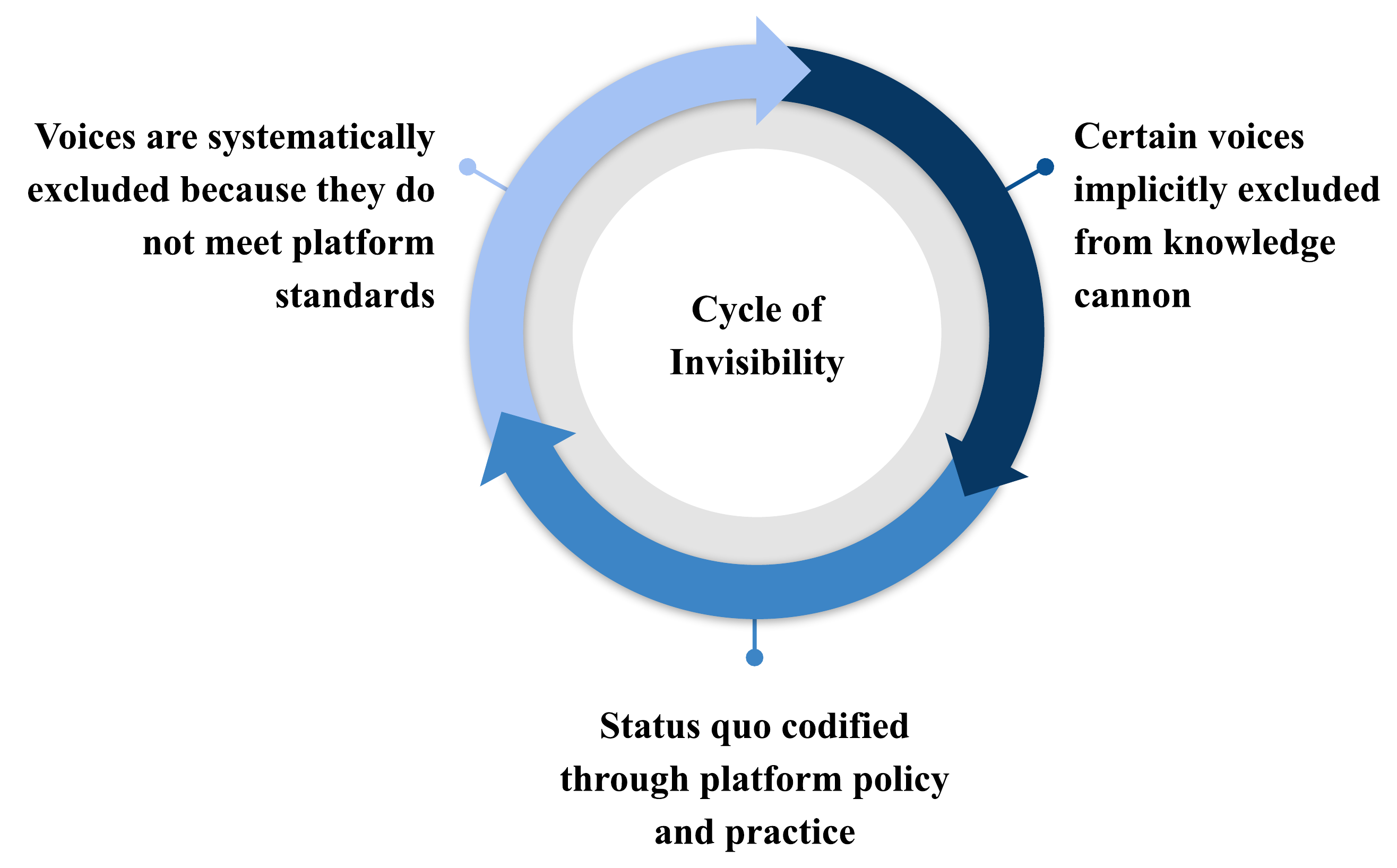}
    \caption{Cycle of invisibility of certain voices.}
    \Description{}
    \label{fig:invisibiilty-cycle}
\end{figure}

The invisibility of relevant perspectives and voices is difficult to detect because of what one participant called \textit{``failures of existing evaluations and benchmarks."} A Wikipedia researcher echoed r/AskHistorian's problems with unfamiliar sources that never become normalized. Noting, 
\begin{quote}
    \textit{``One of Wikipedia's core policies is around Verifiability~\footnote{https://en.wikipedia.org/wiki/Wikipedia:Verifiability}, which, practically speaking, rests on the concept of a reliable source. Whether a source is reliable is a matter of consensus within the editing community but a lot of it relies on familiarity with sources (to be able to judge them). This means that sources that are unknown to the core community can be rejected as non-reliable."}
\end{quote}

CSCW has a rich history of studying the systemic exclusions of certain editors and topics from Wikipedia~\cite{Houtti2022-uo, Menking2019-ep, Lam2011-up, Hargittai2015-yk}. Recent work from~\citet{Menking2021-ic} has noted that Wikipedia's epistemology furthers exclusionary issues on the platform, such as the gender edit gap and lack of notable female article subjects. We describe this general phenomenon across epistemic platforms, such as r/AskHistorians and Wikipedia, as a vicious \textbf{cycle of invisi
bility}. A barrier to collective knowledge is the procedures and processes that keep individuals separate from another story that would otherwise affirm theirs. For example, the removal of submissions in r/AskHistorians codifies a norm of not presenting Indigenous knowledge. When stories are not presented in their true frequency and with all of their composite parts, it presents an inaccurate state of affairs.

The invisibility of reliable but non-traditional sources begets further invisibility, causing a vicious cycle that goes unchecked. As Figure~\ref{fig:invisibiilty-cycle} depicts, once exclusion becomes codified through policy or practice, it becomes more justified to further push those voices out. The cyclical relationship between invisibility and epistemic injustice suggests that information-seeking communities, such as Wikipedia and r/AskHistorians, need to view their epistemic stances as dynamic positions rather than static policies. The notion of a reliable source cannot be a one-time stipulation within a community. Rather, as r/AskHistorians suggests, it is an ongoing conversation with those whose voices have been systemically excluded. Our participants noted promising futures of participatory governance designs---such as digital juries~\cite{Fan2020-pl} or knowledge gap audits~\cite{Wikimedia_Research2022-mt}---that seek to codify more inclusive forms of knowledge.

However, workshop participants noted that visibility is a double-edged sword in CSCW applications, which are ripe with online harassment and toxic hate speech~\cite{Scheuerman2021-nj}. In the case of trans healthcare and r/bisexual, having publically accessible communities creates targets for, at best, tone-deaf discourse and, at worst, unsafe online spaces. For example, women Wikipedians must engage in sophisticated emotional work to feel safe~\cite{Menking2019-ep}. Our work suggests that giving marginalized folks online communities can serve as a collective hermeneutical resource: a space where they can share stories and interpret their own experiences. However, researchers in content moderation and online community health must do the work to ensure that these communities remain safe spaces rather than venues of further harassment.


\subsection{Expertise} In her original work,~\citet{Fricker2007-zh} prefers the term ``knowers" to ``experts" as it distances credibility from large institutions mired in bureaucracy. However, popular media in the early 2000s questioned whether the internet has created a place where ``suddenly, everyone is an expert."~\footnote{https://www.nytimes.com/2000/02/03/technology/suddenly-everybody-s-an-expert.html} Conceptualizing expertise is particularly important for CSCW research, which has continually explored both (1) the harm of rampant misinformation from non-experts~\cite{Efstratiou2022-zh} and (2) the power of involving non-experts into crowd-sourced science~\cite{Pandey2021-qu}. However, our workshop participants found that asking \textit{what defines an expert?} demands simultaneously asking \textit{what is the normative position behind defining expertise?}

If created by the community, traditional expertise can help form strong support systems for said community. As noted in our case on trans healthcare, the Transgender Professional Association for Transgender Health (TPATH)~\footnote{https://tpathealth.org/} is an example of traditional experts (e.g., people with medical degrees) creating a community-based knowledge institution. Concisely, TPATH's mission is ``Trans Health by Trans People." Yet, TPATH has had trouble gaining credibility as a collective of medical health experts, particularly in online spaces with algorithmic moderation. As one participant put it, ``\textit{in online spaces, bodies to formalize knowledge on our own terms are ignored because they are \textit{of} the community, despite the presence of traditional experts.}"

Furthermore, epistemic injustice heavily relies on the continuous interrogation of non-traditional expertise: can someone be a valuable knower without carrying common credentials? This distinction of appropriate level of expertise is especially important when discussing an individual's expertise in their own lived experience. In the case of trans healthcare, individual experiences are valuable information. Yet, they are often sidelined, even in informal online spaces. Some communities have begun to push back against this. For example, in 2020 and 2021, r/AskHistorians moderators organized two academic conferences to be inclusive of lay-historians and democratize scholarly exchange \cite{raeburn2022out}. However, efforts such as these are labor-intensive and rare.

The tension between traditional expertise and owning one's lived experience is echoed in current CSCW research on technology-mediated mental healthcare~\cite{Pendse2021-yu}. Previous work noted that an over-reliance on clinical mental health information creates colonialist narratives~\cite{Pendse2022-mc}. When people are stripped of expertise in their own mental health experience, they are stripped of resources to create their own narratives about their mental illness~\cite{Pendse2023-mb}.~\edit{In other words, they self-discount their own testimony and experience, often preventing them from seeking proper care and community.} Epistemic injustice demands that we explore the potential harm of systemically discounting individuals as experts. In some cases, such as TPATH, we may need to interrogate how our identity biases cloud our understanding of traditional expertise. In lived experiences, such as mental health contexts, we may need to deconstruct the notion of ``expertise'' altogether and allow clinical information to harmonize with anecdotal evidence.

\subsection{Anti-Agentic Discourse}
Finally, workshop contributors reflected inward on discourse within their research communities (CSCW, CHI, FAccT, SIGIR, etc.). Participants noted numerous ways in which language typically associated with raising marginalized identities gets inappropriately wielded, particularly in settings that involve minors, such as trans health or research. We call this \textbf{anti-agentic discourse:} the use of language to discount groups rather than participate with them. Anti-agentic discourse is a form of hermeneutical injustice as it misappropriates an interpretive resource (i.e., language). For example, one participant noted that ``at-risk,'' a term originally used to create more thorough consent procedures in research~\cite{United_States_National_Commission_for_the_Protection_of_Human_Subjects_of_Biomedical_and_Behavioral_Research1979-hg}, \textit{"is thrown around paternalistically in HCI with an exorbitant number of identity groups such as youth, queer folks, substance use, and mental health."} A researcher in online mental health communities echoed this sentiment that protective terms, such as ``safe," have not evolved for modern contexts: 
\begin{quote}
\textit{``We need to loosen the idea of what is `normal', `correct', or `unsafe' when regarding online mental health communities."}
\end{quote}
They suggested contextual, community-specific approaches to defining these terms in research.

Anti-agentic discourse is particularly problematic in research settings. Researchers are canonically accepted as ``knowers:'' people who have the qualities necessary to contribute knowledge. Often in venues such as CSCW and CHI, we are studying marginalized communities who have deep histories of facing epistemic injustice. Therefore, we, as researchers, must engage with the verbiage of these communities rather than the language that we have passed down through archival research papers.

Self-reflection on the implicit positionalities of researching a historically marginalized community is a promising path forward. To paraphrase~\citet{Braun2006-eg}, researchers cannot simply give voice to their participants. As~\citet{Fine1992-ov} argues, ``even a ‘giving voice’ approach involves carving out unacknowledged pieces of narrative evidence that we select, edit, and deploy to border our arguments.''

\section{Epistemic Injustice as a Tool for Researchers and Practitioners}
\begin{table}[H]
\centering
\resizebox{\textwidth}{!}{ 
\begin{tabular}{|l|ccc|cc|c|}
\hline
 &
  \multicolumn{3}{c|}{\cellcolor[HTML]{EFEFEF}\textbf{Practitioner Testimonies}} &
  \multicolumn{2}{c|}{\cellcolor[HTML]{EFEFEF}\textbf{Previous CSCW Work}} &
  \multicolumn{1}{l|}{\cellcolor[HTML]{EFEFEF}\textbf{Personal Experiences}} \\ \cline{2-7} 
 &
  \multicolumn{1}{l|}{Trans Health Care} &
  \multicolumn{1}{l|}{r/AH} &
  \multicolumn{1}{l|}{r/bisexual} &
  \multicolumn{1}{l|}{Wikipedia} &
  \multicolumn{1}{l|}{MH Care} &
  \multicolumn{1}{l|}{Systems of Research} \\ \hline
Invisibility         & \multicolumn{1}{c|}{}  & \multicolumn{1}{c|}{X} & X & \multicolumn{1}{c|}{X} &   &   \\ \hline
Expertise              & \multicolumn{1}{c|}{X} & \multicolumn{1}{c|}{X} & X & \multicolumn{1}{c|}{}  & X & X \\ \hline
Anti-Agentic Discourse & \multicolumn{1}{c|}{X} & \multicolumn{1}{c|}{}  &   & \multicolumn{1}{c|}{}  & X & X \\ \hline
\end{tabular}}
\caption{Common tensions among cases from our collaborative autoethnography ($n$ = 14). r/AskHistorians abbreviated to r/AH. Online mental health care abbreviated to MH Care.}
\label{tab:theme-overview}
\end{table}

CSCW is rapidly growing, expanding from less than 50 articles per-year in 1990 to over 200 articles per-year in 2023. Moreover, CSCW researchers gather together annually, participate in town halls, and need to have a modern understanding of the field for their own research. In other words, researchers are encouraged to create implicit understandings of the field. As the discourse on epistemic concerns within HCI and CSCW grows~\cite{Pierre2021-sv, Collective2021-cg}, there is a simultaneous burden put on members of our community to speak out and remediate this injustice themselves~\cite{Erete2021-xp}. Can we support this work without collectively placing labor on our already marginalized colleagues?

Our workshop found that epistemic injustice is banal and difficult to measure and, therefore, feeds off of problematic implicit understandings. Making the status quo explicit is a key to understanding where CSCW is failing certain people. For example, r/AskHistorians realized they had a systematic gap of Indigenous knowledge once they reflected on their Western-centric model of ``knowledge.'' Members of r/bisexual had not fully recognized their lack of hermeneutical resources until they found others with similar experiences. Historically, concepts that give voice to injustice towards women, such as sexual assault and post-partum depression, went misunderstood and unarticulated~\cite{Fricker2007-zh}. Our work suggests that CSCW risks perpetrating these epistemic injustices if it relies solely on implicit understandings.

CSCW needs an explicit understanding of the field and the specific knowledge gaps that go unnoticed. Many practitioners have called for a larger synthesis of CSCW as a field through systematic literature reviews that surface ``sticky'' practices~\cite{Ajmani2023-bg} and reflexive work~\cite{Wallace2017-mi}. Others have advocated for a greater emphasis on history in CSCW to combat our field's presentism \cite{soden2021time}. We raise the stakes of this problem area: not understanding the knowledge gaps of CSCW propagates injustice. When we fail to interrogate the voices, stories, and values that CSCW excludes, we fail to do right by the people who make our research possible. Can we synthesize the epistemology of contemporary CSCW work as a means towards justice? Can we stop cycles of epistemic injustice by encouraging more reflexive research on the types of problems, narratives, and populations that CSCW studies? We posit that the various CSCW research agendas surrounding truth and knowledge must be unified under epistemic injustice to properly situate the systemic harm at stake.

\subsection{Epistemic Injustice as a Unifying Lens}
Epistemic injustice surfaces in numerous CSCW research areas, such as content moderation, online peer support, and peer-production. We suggest that epistemic injustice can unify these research agendas under a single umbrella with the goal of supporting equitable contribution and access to online knowledge.

Our workshop surfaced examples that echo previous CSCW research problems (see Table~\ref{tab:theme-overview}), but little work has explored the underlying normative stance that connects these. Some of these connections are obvious, for example, r/AskHistorians and Wikipedia have similar information seeking goals and rely on user-generated content. However, Reddit is often viewed as a social media site in CSCW research~\cite{Proferes2021-tu, Reagle2022-yx} rather than a peer-production site that creates knowledge. When we view both platforms as \textit{epistemic communities}---self-governed online spaces that seek to create and curate knowledge---we can properly articulate the potential injustices that unchecked positions propagate. 

Viewing seemingly disparate platforms, such as social media sites and peer-production, as epistemic communities is especially important when considering them as online resources for marginalized populations. As our exploration of r/bisexual suggests, even subreddits that are not explicitly public knowledge communities are epistemic resources. For example, r/bisexual documents the nuance behind using labels. Therefore, the community provides hermeneutical resources that help remediate bi-erasure and stigmatization. Previous work has found that mental health discourse on TikTok is a crucial resource for support-seeking behavior~\cite{Milton2023-vu}. However, TikTok uses its own ``algospeak'' to create a whole new vocabulary for describing mental health experiences without triggering automoderators~\cite{Steen2023-gb}. Epistemic injustice suggests that unfettered moderation of this ``algospeak'' may strip the mental health community of necessary interpretative resources. Therefore, CSCW research must first understand the community and knowledge dynamics of TikTok before exploring content moderation on the platform.

As information dynamics online span peer-production, social media, search engines, and even large-language models, it is crucial that CSCW finds the interactions across various research agendas. Epistemic injustice gives us a vocabulary to describe why a subreddit might mirror the dynamics on Wikipedia and how exclusion on one platform, such as r/AskHistorians, may signal pervasive exclusion of voices, such as Indigenous voices, from knowledge contribution as a whole.

\subsection{Epistemic Injustice as an Intersectional Lens}
Finally, we suggest that epistemic injustice itself is a hermeneutical resource for marginalized researchers within CSCW: a way of interpreting conflicting and overlapping positionalities. For example, as researchers, we choose whose stories merit inquiry. Simultaneously, many of us have had our own stories silenced both in CSCW applications, such as the over-moderation of trans folks, and in societal ways, such as not being believed during police testimony. We believe that it is important for CSCW to embrace and actively navigate this signature tension of the field in the same way we navigate the tension between ``social" and ``technical."

Our workshop was full of paradoxical observations of epistemic injustice as CSCW practitioners have intersecting identities as (1) credible keepers of the ``knowledge cannon'' of research and (2) marginalized folks moving through the world. We posit that these paradoxes are an instantiation of intersectionality: the philosophy that various identities---such as gender, race, and societal position---are not discrete categories of an individual~\cite{Collins2020-bx}. Rather, identity characteristics interact with one another to create complex portraits of power and marginalization. In exploring Intersectionality for HCI,~\citet{Schlesinger2017-af} recommend providing author disclosures to, ``\textit{clarify who the we each paper mentions is and how that influences the research.}" However, reflexive considerations are a daunting task as they force us to surface the relevant parts of our identity and power as a researcher~\cite{Liang2021-qb}.

Epistemic injustice gives researchers the tools to understand their own power within systems of research. Similar to race, gender, sexuality, and other identity characteristics, we suggest that \textit{epistemic power}---the likelihood of one's testimony to be believed as knowledge---is an intersecting identity that all CSCW practitioners hold. This idea of epistemic power allows us to reconsider our research methodologies and biases. When we report results, are we ignoring narratives that we disagree with? How can we choose measures that properly quantify the systemic exclusion of voices from CSCW platforms? How can we interpret the stories we hear without imposing power onto them? While research is a powerful tool to surface marginalized voices, it also affords the researcher epistemic power. We suggest that this epistemic power is a necessary intersecting identity for all CSCW practitioners to consider during their own work. 

\section{Calls to Action for CSCW}
In this section, we outline the actionable steps for CSCW to move towards epistemic justice. There is enormous potential for CSCW practitioners to leverage the vocabulary of epistemic injustice to make CSCW more collaborative, inclusive, and, ultimately, a more just venue for knowledge.

\edit{\textbf{Centering Individual Experience Through Testimonial Methods. }}
\edit{The action towards centering individuals may at first appear counter to the ``cooperative work'' of CSCW, but it is actually fundamental. As our testimonies in this paper demonstrate, the individual is the most granular level at which people experience epistemic injustice.
Acknowledging and responding to this reality requires flexibility in the formats available to make those individual experiences visible and comprehensively illustrated. As we have illustrated here with testimonies, each individual story is rich with context, perspective, and direction and warrants sufficient space for full expression. A demonstrative step towards centering individual experience is for CSCW to create a track where testimonial methods, such as autoethnography and storytelling, are celebrated. This call echoes similar imperatives for racial justice in HCI~\cite{Ogbonnaya-Ogburu2020-wt, Erete2021-xp, Schlesinger2017-af}. The creation of this track would not only confront the power dynamics within the curation of epistemic contributions, but it would also create a specific space for longer form submissions that contain these important experiences}.

\edit{\textbf{Bolstering Communities Through Our Research Goals.}}
\edit{If epistemic injustice is surfaced in individual testimony, it is resisted through communal discourse. Our work suggests that the collection of individual experiences facilitates resistance to the material harms of epistemic injustice---such as isolation, fragmentation, and marginalization. This is illustrated in Fricker's example of ascertaining the term ``sexual harassment'' through consciousness-raising~\cite{Hogeland2016-gj}. }

Our work highlights that online communities are similarly building a collective resistance to epistemic injustice. As a field that prides itself on building ethical systems, can we express solidarity with these communities, such as r/bisexual, that are using technology to provide necessary epistemic resources? We suggest that CSCW requires a paradigm shift towards research as solidarity with our sites of study. Once we understand the needs of the community, can we build systems and theories that directly benefit them rather than writing a research paper and walking away?~\edit{Particularly when studying communities who are in active fights for basic rights, can we design within the fraught political context they live in everyday~\cite{Asad2019-hd}?} In the spirit of research as solidarity, the authors of this work have committed to sustained efforts of surfacing epistemic injustice within CSCW work. For example, we plan to continue running researcher workshops to build a repository of autoethnographic materials. We plan to disseminate educational resources, such as reading lists and lesson plans, in collaboration with related advocacy groups, such as Queer in HCI~\cite{DeVito2021-pw} and the WikiWomen's user group \footnote{ \protect\url{https://meta.wikimedia.org/wiki/WikiWomen\%27s\_User\_Group} }.

\edit{\textbf{Remeditating Issues of Power by Answering Epistemic Questions.}}
Recall that epistemology refers to the underlying presumptions we make about how we form knowledge as distinct from belief. In research, epistemology sits in the space between our understanding of what exists our actionable ways of acquiring the knowledge we seek. CSCW has a diverse methodology, ranging from purely quantitative to theoretical~\cite{Oulasvirta2016-en}, but much of CSCW discussion has stayed there. Our work suggests that undiscussed epistemic stances carry dramatic consequences for how we treat the communities we try to serve.~\edit{In research, we have epistemic power over our sites of study.} For example, our protective terminology such as ``at-risk'' has since been weaponized to propagate paternalistic research practices. We suggest that CSCW leverage its pre-existing reflexive venues, such as town halls~\cite{Bruckman2017-qq, Munteanu2019-az}, workshops~\cite{Ajmani2023-cp}, panels \cite{ribes2023history}, and reflexive papers~\cite{Wallace2017-mi, Erete2021-xp} to discuss larger epistemic questions. For example, what does it mean to “give voice” to research participants? Are we, as CSCW practitioners with our own intersecting identities, coming in as outsiders and mirroring colonialist practices of exerting power through selective narratives~\cite{Schlesinger2017-af, Pendse2022-mc}? Even if we are member researchers, are we appropriately recognizing the epistemic power we hold or letting our own biases paint skewed portraits of already marginalized communities? Who do our conventional methodologies systemically exclude~\cite{harrington2019deconstructing}? Our work suggests that CSCW needs to open up a variety of venues to have these rich discussions and eventually arrive at a set of epistemic standards for the field.


To summarize, CSCW can leverage many characteristics of the field, such as its reflexive town halls and research focus on justice, to operationalize epistemic justice. CSCW is equipped to move toward epistemic justice by doing more of what comes naturally: participatory and collaborative practices that \textit{surface} individual articulations of experience, designing to \textit{support} communities, and \textit{signal} to issues of power and positionality in the execution of both. Thus, the use of epistemic injustice as a theoretical and analytic framework facilitates lateral steps in these areas to \textit{center} individual experience, \textit{bolster} communities, and \textit{remediate} issues of power. We see a future for the field to become a paragon of epistemic justice that can serve as an example for related research communities. The authors of this work are committed to engaging in sustained efforts so that this just future can be realized in our own research community.

\section{Limitations}
We chose to use collaborative autoethnography methods for this work so we could highlight personal experiences and common tensions of epistemic injustice. One limitation of this method is that it does not comprehensively capture epistemic injustice. As~\citet{Fricker2007-zh} articulates, epistemic injustice is ubiquitous yet normalized. Therefore, it surfaces in many settings that we did not mention in this paper. Future work could use larger-scale methods, such as surveys, to gain a broader picture of epistemic injustice and its scope.

Moreover, our own positionalities as solution-oriented researchers may undersell the nuance of investigating epistemic injustice. We offer mechanisms for preventing and repairing epistemic injustice, but we also urge researchers to focus on understanding the community they research before imposing solutions. As~\citet{Baumer2011-or} articulate, sometimes the implication is not to design.

\section{Conclusion}
In this paper, we have contributed a new understanding of how sociotechnical settings can further injustice through their ways of knowing. We present three cases of epistemic injustice in sociotechnical applications: online trans healthcare, identity sensemaking on r/bisexual, and indigenous ways of knowing on r/AskHistorians. We further explore common tensions across our autoethnographic materials and relate them to previous CSCW research areas and personal non-technological experiences. We argue that epistemic injustice can serve as a unifying and intersectional lens for CSCW research by reframing certain online spaces as epistemic communities. It is our hope that this work represents a recommitment to studying the normative consequences that come with epistemic commitments. We offer calls to action for the CSCW community and hope to spark a larger dialogue on furthering epistemic justice.



\bibliographystyle{ACM-Reference-Format}
\bibliography{main}


\end{document}